%% file: main.tex
\pdfoutput=1
\documentclass{llncs}
\usepackage{esvect}
\usepackage{graphicx} 
\usepackage[T1]{fontenc}
\usepackage{amsmath,amssymb}
\usepackage{booktabs}
\usepackage{array}
\usepackage{tabularx}
\usepackage{enumitem}
\usepackage{xcolor}
\usepackage{hyperref}
\usepackage[sorting=none]{biblatex} 
\usepackage{tikz}
\usetikzlibrary{arrows.meta,positioning}
\usepackage{url}
\usepackage{xspace}
\hypersetup{hidelinks}

\newcommand{\secodd}{\textsc{SecODD}\xspace}
\newcommand{\msh}{\textsc{MSH}\xspace}
\newcommand{\rade}{\textsc{R-A-D-E}\xspace}
\newcommand{\ind}{\mathbb{1}}

\title{SecODD: Adapting the Operational Design Domain from Driving Autonomy to Autonomous Cyber Defense}

\author{Jina Kang\inst{1} \and Jeong-Han Yun\inst{1} \and Hyesong Choi\inst{2}\thanks{Corresponding author.}}
\authorrunning{J. Kang et al.}
\institute{National Security Research Institute, Daejeon, Republic of Korea
\and
School of Electronic Engineering, Soongsil University, Seoul, Republic of Korea}

\begin{document}

\maketitle

\input{sec/0_abstract_4}

\input{sec/1_introduction_4}
\input{sec/2_background_6}
\input{sec/3_framework_5}
\input{sec/4_normative_5}
\input{sec/5_engreq_3}
\input{sec/6_scenarios_4}
\input{sec/7_conclusion_3}

\printbibliography

\end{document}

%% file: sec/0_abstract_4.tex
\begin{abstract}

Security operations centers are beginning to delegate detection, triage, and containment to
Autonomous Cyber Defense (ACD) agents at execution time. Erroneous high-impact playbooks can
disrupt critical services and create accountability gaps among deployers, SOC supervisors,
vendors, managed service providers, and infrastructure providers.
We propose \textsc{SecODD}, an execution-time governance framework for ACD that adapts the
Operational Design Domain concept from autonomous-vehicle regulation to cyber autonomy. It
defines the authorized scope of automated defensive actions, models when meaningful human
intervention must remain possible, and specifies auditable due-care conditions through a
Mitigation Safe Harbor. Four engineering requirements---tamper-evident logging, human-in-the-loop
orchestration, rollback playbooks, and safe-state transition controls---are evaluated through a per-scenario audit matrix that illustrates how two ACD
failure modes can be translated into pass/fail audit outcomes. 
The result is an evidence-oriented framework that authorizes ACD only when autonomous actions are bounded,
interruptible, recoverable, and reconstructable from evidence.

\keywords{Autonomous Cyber Defense \and Security Operations Center \and AI Governance \and
Operational Design Domain \and Safe Harbor \and Human-in-the-Loop \and Incident Response}
\end{abstract}

%% file: sec/1_introduction_4.tex
\section{Introduction}
\label{sec:intro}

Generative-AI tooling is narrowing the time gap between cyber discovery, exploitation, and
response, with defensive systems increasingly supporting alert triage, correlation, and playbook
execution~\cite{acd-isa2023,lee2025supplychainai,kim2025llmguardrail};
fully unattended cyber defense is not yet ordinary practice, but bounded human-machine teaming is
becoming operationally relevant~\cite{cetas2024acd}.

This shift creates an \emph{execution-time governance} problem. When an Autonomous Cyber Defense
(ACD) agent misclassifies a benign workload and triggers a high-impact playbook, responsibility
may involve the deployer, SOC supervisor, AI vendor, playbook supplier, MSSP, cloud provider, and
identity provider. Existing AI-governance and cybersecurity-risk frameworks do not answer the
action-boundary questions at the moment of execution: whether the action was inside the
authorized envelope, whether a qualified human could have intervened in time, whether rollback or
safe-state transition was rehearsed, and whether evidence is sufficient to reconstruct
responsibility. 
Recent work on zero-trust guardrails and generative-AI supply-chain security
shows that governance mechanisms are more useful to AI-enabled security systems
when they are grounded in auditable engineering controls~\cite{kim2025llmguardrail,lee2025supplychainai}.

This paper proposes \secodd, an execution-time governance framework for ACD that ports reusable
primitives from autonomous-vehicle governance---bounded operating domains, accountable roles,
override or handover, event evidence, and residual loss allocation---and reifies them as a
Security ODD, a control-state set $\Omega$, an intervention-probability function
$P_{\mathrm{int}}$, a Role-Authority-Duty-Evidence (\rade) model, and a Mitigation Safe Harbor
(\msh). The paper makes four contributions.

\begin{description}[leftmargin=1.4em,itemsep=0pt,topsep=1pt,parsep=0pt]
\item[\textbf{C1.}] \secodd, a formal framework with a four-dimensional Security ODD, a discrete
control-state set $\Omega$, a measurable intervention-probability function $P_{\mathrm{int}}$, and
an interpretable apportionment function $L$ (Section~\ref{sec:framework}).
\item[\textbf{C2.}] An evidence-based accountability model with a three-tier autonomy classification, a
\rade tuple, and a seven-condition \msh that maps due-care
claims to auditable artifacts across NIST, ISO, and OECD-style governance
regimes (Section~\ref{sec:normative}).
\item[\textbf{C3.}] Four audit-grade engineering requirements that make \msh auditable:
tamper-evident logging via a one-way evidence channel, HITL orchestration, rollback playbooks,
and multi-channel safe-state transition, with AIBOM-linked traceability
(Section~\ref{sec:engreq}).
\item[\textbf{C4.}] A scenario-based evaluation of two ACD failure scenarios, with illustrative
pass/fail audit outcomes and a Korean operational-profile mapping that identifies
implementation gaps and a three-layer roadmap (Section~\ref{sec:scenarios}).
\end{description}

\noindent \secodd reframes the question from \emph{whether} ACD should be allowed in the abstract
to \emph{when} a specific mitigation action is bounded, interruptible, recoverable, and
reconstructable from evidence.

%% file: sec/2_background_6.tex
\section{Background: Governance Gaps in Autonomous Cyber Defense}
\label{sec:background}

\secodd rests on a narrow premise: the hard governance problem in ACD arises at \emph{execution
time}, when an SOC agent recommends, sequences, or executes a playbook that may cross a
service-critical boundary. Existing instruments operate \emph{upstream}: AI-governance
frameworks address risk management, accountability, and incident
comparability~\cite{nist-airmf,nist-ai-profile2025,iso42001,oecd-ai-incidents,eu-ai-act}, while
cybersecurity standards cover controls, incident response, logging, resilience, and defensive
mapping~\cite{nist-csf20,nist-80061r3,nist-80092,nist800160v2r1,mitre-attack,mitre-d3fend}.
Supply-chain provenance schemes improve artifact
traceability~\cite{cisa-sbom-min-elements,spdx-ai-profile,slsa,intoto}, and
product-security regimes motivate evidence and reporting
discipline~\cite{eu-cra2024,eu-nis2}.

None of these specifies the \emph{last mile} of ACD execution: whether a containment action lies
inside the authorized operating envelope, whether a qualified human could intervene in time,
what counts as evidence of due care, and how residual loss is apportioned once a high-impact
playbook has fired. ACD additionally requires \emph{action-time} traceability binding each
executed command to the model, prompt, policy, plugin, and playbook version that produced it.
Autonomous-vehicle (AV) governance offers a structural analogy: it already combines five
primitives that ACD lacks at the execution boundary---bounded operating domains, accountable
entities, override or handover, event evidence, and residual loss
allocation~\cite{uk-ava2024,uk-aeva2018,germany-av-act,eu-pld2024,eu-av-typeapproval2022,japan-level4-guide}.
\secodd reuses these as a Security ODD, actor lattice, control-state set, tamper-evident
evidence channel, and MSH-based apportionment, while AIBOM-linked traceability and safe-state
transition controls (Section~\ref{sec:engreq}) address ACD's adversarial, multi-tenant
substrate.

%% file: sec/3_framework_5.tex
\section{SecODD Framework for Security Autonomy}
\label{sec:framework}

\secodd translates the AV governance primitives of Section~\ref{sec:background} into a
cyber-autonomy framework---actor mapping, Security ODD, control states, intervention probability,
cyber-specific limits, and apportionment---supplying the vocabulary used by \rade and \msh
(Section~\ref{sec:normative}), the engineering requirements (Section~\ref{sec:engreq}), and the
scenario walks (Section~\ref{sec:scenarios}).

\subsection{Actor Mapping}
\label{sec:actors}

The AV \emph{driver} does not map onto a single SOC actor. Operational authority is split
between the \emph{Deploying Organization}, which authorizes the agent, and the \emph{SOC
Supervisor}, who occupies the functional human-supervision role during automated operation;
system responsibility is distributed across AI vendors, playbook or model suppliers, MSSP/MDR
providers, and infrastructure providers. Table~\ref{tab:actors} gives a compact actor lattice;
Section~\ref{sec:rade} expands it into the \rade tuple.

\begin{table}[tbp]
\centering
\caption{Compact actor lattice for porting AV governance to ACD.}
\label{tab:actors}
\scriptsize
\setlength{\tabcolsep}{0pt}
\renewcommand{\arraystretch}{1.04}
\begin{tabular}{@{}
>{\raggedright\arraybackslash}p{0.28\linewidth}
@{\hspace{3pt}}
>{\raggedright\arraybackslash}p{0.36\linewidth}
@{\hspace{3pt}}
>{\raggedright\arraybackslash}p{0.30\linewidth}
@{}}
\toprule
\textbf{Functional analogue} &
\textbf{ACD counterpart} &
\textbf{\secodd duty anchor} \\
\midrule
Human authority &
Deploying organization; SOC supervisor &
ODD ratification; HITL readiness \\
System supply and operation &
AI security vendor; playbook/model supplier; MSSP/MDR &
Envelope validation; safety updates; monitoring \\
Infrastructure and allocation &
Cloud/identity provider; insurer; competent authority &
Evidence delivery; reconstruction; certification/registry \\
\bottomrule
\end{tabular}
\end{table}

\subsection{Security ODD: Definition and Worked Example}
\label{sec:secodd-def}

The AV Operational Design Domain becomes a cyber safety envelope over protected assets, temporal
windows, data-sensitivity levels, and permitted defensive responses.

\begin{definition}[Security ODD]
\label{def:secodd}
Let $\mathcal{A}$ be \emph{asset classes} (e.g., workloads, identities, network segments, and data tiers);
$\mathcal{T}$ \emph{temporal windows} (e.g., supervised operation, approved low-change windows, maintenance, and change-freeze periods);
$\mathcal{D}$ \emph{data-sensitivity levels} (e.g., ISMS-P); and 

$\mathcal{P}$ \emph{permitted defensive responses} (e.g., observe, alert, throttle, isolate, revoke, terminate, and modify policy). 
The \emph{Security ODD} authorized for an autonomous agent is
\begin{equation}
\mathrm{ODD}_{\mathrm{sec}}
\;\subseteq\;
\mathcal{A} \times \mathcal{T} \times \mathcal{D} \times \mathcal{P},
\label{eq:secodd}
\end{equation}
with an extension predicate $\Phi(\omega,c)$ that admits scenario-conditioned regimes
$\omega \in \mathrm{ODD}_{\mathrm{sec}}$ 
under observable context $c$ (e.g., incident severity,
supply-chain trust state, and attestation freshness).
\end{definition}

\noindent Table~\ref{tab:oddexample} instantiates Definition~\ref{def:secodd} for an enterprise SOC. 
The predicate $\Phi$ keeps the Security ODD from collapsing into a static checklist: a high-severity ransomware indicator combined with a fresh attestation may temporarily admit production-asset isolation that the static Security ODD would otherwise forbid.

\begin{table}[tbp]
\centering
\caption{Worked example: a Security ODD declared by an enterprise SOC. Each row instantiates one
dimension of Eq.~\eqref{eq:secodd}.}
\label{tab:oddexample}
\scriptsize
\setlength{\tabcolsep}{0pt}
\renewcommand{\arraystretch}{1.04}
\begin{tabular}{@{}
>{\raggedright\arraybackslash}p{0.10\linewidth}
@{\hspace{3pt}}
>{\raggedright\arraybackslash}p{0.27\linewidth}
@{\hspace{3pt}}
>{\raggedright\arraybackslash}p{0.24\linewidth}
@{\hspace{3pt}}
>{\raggedright\arraybackslash}p{0.33\linewidth}
@{}}
\toprule
\textbf{Dim.} & \textbf{Authorized values} & \textbf{Forbidden values} & \textbf{Effect on $\mathcal{P}$} \\
\midrule
$\mathcal{A}$ &
Non-production endpoints; sandboxed identities &
Production IAM roles; customer-facing services &
Isolate / revoke permitted on the left; dual approval required on the right. \\
$\mathcal{T}$ &
Supervised normal operation; approved low-change windows & 
Maintenance, migration, deployment, and change-freeze windows &
Auto-containment falls back to $s_0$ when change context is missing or stale. \\
$\mathcal{D}$ &
Public, internal, ISMS-P low &
Sensitive personal data; trade secrets &
Sessions touching sensitive data require explicit pre-authorization. \\
$\mathcal{P}$ &
Observe, alert, throttle, host-level isolate &
Tenant-wide credential revocation; cross-tenant policy mutation &
Forbidden actions trigger an \msh-revocation event (Section~\ref{sec:msh}). \\
\bottomrule
\end{tabular}
\end{table}

\subsection{Control-State Set and Intervention Probability}
\label{sec:omega}

\begin{definition}[Control-state set]
\label{def:omega}
Let $\Omega=\{s_0,s_1,s_2,s_3\}$: $s_0$ \emph{Advisory} (agent proposes, human ratifies;
e.g., a SOAR playbook awaiting analyst sign-off); $s_1$ \emph{Reversible Automated} (rollback
within budget $\tau_R$; e.g., host-level quarantine reversible within minutes); $s_2$
\emph{High-Impact Automated} (pre-authorized with mandatory post-hoc review; e.g.,
service-principal freeze on a critical workload); $s_3$ \emph{Halt} (agent suspended and
authority reduced to a minimum-risk surface).
\end{definition}

A transition function $\delta:\Omega\times\mathcal{E}\to\Omega$ governs state changes under
alarms, transition demands, overrides, safe-state invocations, and update-integrity failures. The
intervention-probability function
\begin{equation}
P_{\mathrm{int}}(s_i,t)
=
\Pr\!\big[
\text{qualified override within } \tau_O
\,\big|\, s_i,t
\big]
\label{eq:pint}
\end{equation}
turns human oversight into an \emph{auditable variable}, estimated from acknowledgement logs,
override-channel availability, and on-call attestation.

\subsection{Cyber-Specific Limits and Apportionment}
\label{sec:Lfn}

The AV analogy has three cyber-specific limits: ACD harm is distributed across multi-tenant,
multi-provider substrates rather than localized at a single physical point; ACD operates over a
transitive software and service supply chain, making AIBOM identifiers first-class \rade evidence
inputs~\cite{spdx-ai-profile}; and the cyber environment is adversarial, so the logging and
guardrail substrate is itself an attack target. These motivate the evidence and control
requirements of Section~\ref{sec:engreq}.

For an incident-time action-context tuple $a_t$, the apportionment intensity at SOC playbook
execution time is
\begin{equation}
L(t,s_i,a_t)
=
\alpha \cdot
\ind\!\big[a_t \notin \mathrm{ODD}_{\mathrm{sec}}\big]
+
\beta \cdot
\big(1-P_{\mathrm{int}}(s_i,t)\big)
+
\gamma \cdot
\ind\!\big[\text{\msh\ violated}\big],
\label{eq:Lfn}
\end{equation}
with policy-set weights $\alpha,\beta,\gamma\ge 0$. $L$ is a \emph{measurement model}, not a
damages function: each term maps to an auditable artifact (the ODD specification, the
acknowledgement log, the \msh evidence pack), so $L$ structures the post-incident evidence walk
rather than computing a monetary outcome. Lower $L$ indicates lower deployer-side accountability intensity for the executed action; residual allocation across other actors is handled through the actor lattice (Table~\ref{tab:actors}) and the \rade evidence walk (Section~\ref{sec:rade}).

%% file: sec/4_normative_5.tex
\section{Evidence-Based Accountability Model}
\label{sec:normative}

This section converts \secodd's vocabulary into a governance model where tiers trigger
obligations, \rade reconstructs responsibility, and \msh defines auditable due care.

\subsection{Three-Tier Autonomy Classification}
\label{sec:tiers}

\secodd partitions ACD deployments into three obligation tiers anchored to $\mathcal{P}$ of
Eq.~\eqref{eq:secodd} and $\Omega$:

\begin{description}[leftmargin=1.35em,itemsep=0pt,topsep=1pt,parsep=0pt]
\item[\textbf{T1 -- Advisory ACD}] ($s_0$): agent enriches, prioritizes, and proposes without
active mitigation; baseline AI governance applies, no \msh required.
\item[\textbf{T2 -- Reversible Automated ACD}] ($s_1$): agent executes rollback-bounded
containment within $\tau_R$; risk assessment, override, logging, rollback, and safety-update
controls are mandatory.
\item[\textbf{T3 -- High-Impact Automated ACD}] ($s_2$): agent executes legally consequential or
business-critical actions (credential revocation, data quarantine, cross-tenant policy mutation);
full \msh, formal authorization, independent validation, financial responsibility, and registry
reporting apply.
\end{description}

\noindent The tier is an \emph{obligation trigger}, not a maturity label, allowing sectors to add
requirements without redefining \secodd.

\subsection{Role-Authority-Duty-Evidence (R-A-D-E) Matrix}
\label{sec:rade}

For each actor $x$ in Table~\ref{tab:actors}, \rade records
\begin{equation}
\mathrm{RADE}(x)=
\langle r_x,\, A_x,\, D_x,\, E_x \rangle ,
\label{eq:rade}
\end{equation}
the role, authority, duty, and evidence. Table~\ref{tab:rade} instantiates Eq.~\eqref{eq:rade}
for three representative actors; the post-incident walk over each row drives the apportionment
in Eq.~\eqref{eq:Lfn}.

\begin{table}[tbp]
\centering
\caption{\rade instantiation for three representative actors. Cells map directly to the evidence
artifacts examined by the post-incident walk in Section~\ref{sec:scenarios}.}
\label{tab:rade}
\scriptsize
\setlength{\tabcolsep}{0pt}
\renewcommand{\arraystretch}{1.04}
\begin{tabular}{@{}
>{\raggedright\arraybackslash}p{0.18\linewidth}
@{\hspace{3pt}}
>{\raggedright\arraybackslash}p{0.21\linewidth}
@{\hspace{3pt}}
>{\raggedright\arraybackslash}p{0.26\linewidth}
@{\hspace{3pt}}
>{\raggedright\arraybackslash}p{0.29\linewidth}
@{}}
\toprule
\textbf{Actor} & \textbf{Authority $A_x$} & \textbf{Duty $D_x$} & \textbf{Evidence $E_x$} \\
\midrule
Deploying organization (SOC supervisor) &
Issue transition demand; invoke safe-state; suspend tier &
Author and version ODD; ratify high-impact actions; rehearse override and rollback &
ODD versions; HITL acknowledgement logs; override-drill records; safe-state invocation logs \\
AI security vendor &
Push safety-critical updates; flag drift; deprecate playbooks &
Validate operating envelope; disclose known limits; meet update SLA &
Model and policy versions; update attestations; red-team and shadow-mode reports; AIBOM IDs \\
MSSP / MDR provider &
Escalate to deployer; trigger registry report; gate executions awaiting approval &
Monitor within agreed SLA; preserve telemetry; cooperate with forensic walk &
Escalation log; SLA-conformance reports; incident attestations; evidence-chain handover \\
\bottomrule
\end{tabular}
\end{table}

\subsection{Mitigation Safe Harbor (MSH)}
\label{sec:msh}

\msh is an engineering-auditable evidence regime inspired by autonomous-vehicle
safe-harbor architectures~\cite{uk-aeva2018}.
A deployment is \msh-qualified only if all seven gates hold:

\begin{enumerate}[
  label=\textbf{MSH-\arabic*.},
  labelwidth=3.8em,
  labelsep=0.6em,
  leftmargin=4.6em,
  align=left,
  itemsep=0pt,
  topsep=2pt,
  parsep=0pt
]
\item \emph{ODD declaration:} versioned $\mathrm{ODD}_{\mathrm{sec}}$ and forbidden-action list.
\item \emph{HITL provisioning:} pre-authorization and bounded review for $s_2$ within $\tau_O$.
\item \emph{Independent evaluation:} red-team and shadow-mode evidence within the declared ODD.
\item \emph{Tamper-evident logging:} inputs, model and policy versions, scores, commands,
overrides, and update state.
\item \emph{Update integrity:} safety-critical updates applied within the vendor deadline.
\item \emph{Override viability:} multi-channel override and rollback rehearsals on a published
cadence.
\item \emph{Incident attestation:} 24h/72h/30d reporting to the competent registry.
\end{enumerate}

\noindent
\msh is not proposed as automatic legal immunity or a statutory safe harbor; it is an evidence structure for assessing whether due-care conditions were satisfied under a declared Security ODD.

In Eq.~\eqref{eq:Lfn}, the \msh-violation term is zero unless the evidence
walk shows material failure of one of these gates.

\subsection{Post-Incident Accountability Allocation and Disclosure Controls}
\label{sec:marketing}

For T2 and T3, \secodd supports post-incident allocation through the \rade evidence trail rather than ex ante blame assignment. The evidence walk identifies which control gate failed and which actor was positioned to prevent or detect it: the deployer for ODD, logging, or update-integrity failures; the vendor for unsafe model, policy, or playbook updates; and the infrastructure provider for evidence-delivery failures. T1 products must not be marketed as ``full autonomy'', ``zero false positives'', or unqualified ``AI security''; claims must disclose tier, declared Security ODD, and \msh status.

%% file: sec/5_engreq_3.tex
\section{Engineering Requirements for Auditability}
\label{sec:engreq}

\secodd's governance model becomes operational through four engineering requirements that make
oversight, evidence, rollback, and safe halt auditable
(Table~\ref{tab:engreq-overview}).

\begin{table}[t]
\centering
\caption{Engineering requirements, measurable checks, and auditable evidence.}
\label{tab:engreq-overview}
\scriptsize
\setlength{\tabcolsep}{2pt}
\begin{tabularx}{\textwidth}{p{0.8cm}XXX}
\toprule
\textbf{Req.} & \textbf{Control objective} & \textbf{Measurable check} & \textbf{Evidence artifact} \tabularnewline
\midrule
R1 &
Preserve execution-time evidence &
EC; hash-chain continuity; timestamp validity &
Raw event; command log; model and policy version; AIBOM ID \tabularnewline
\midrule
R2 &
Keep human intervention meaningful &
p95 OLA; OVS; dual approval for T3 &
Approval log; dashboard acknowledgement; override-channel test \tabularnewline
\midrule
R3 &
Ensure rollback or minimum-risk recovery &
Rollback pass rate; RCT &
Rehearsal record; restoration timestamp; fallback plan \tabularnewline
\midrule
R4 &
Enable safe halt and authority reduction &
Dual-channel halt success; authority reduction &
Safe-state log; break-glass record; dependency map \tabularnewline
\bottomrule
\end{tabularx}
\end{table}

\paragraph{R1 -- Tamper-evident logging.}
\emph{Engineering claim:} every execution-time decision must be reconstructable from a vault the
agent cannot rewrite. The telemetry channel captures input alerts, context signals, model and
policy versions, confidence scores, action candidates, executed commands, override events, update
state, and AIBOM identifiers; for high-impact deployments, \secodd recommends a one-way
evidence channel or data-diode-class capture path. \emph{Evidence Completeness} (EC) measures the
fraction of required artifacts available at incident time.

\paragraph{R2 -- Human-in-the-loop orchestration.}
\emph{Engineering claim:} a dashboard is not oversight unless a qualified human can both
understand and override the action within $\tau_O$. The interface exposes action candidates,
affected assets, ODD status, uncertainty signals, playbook consequences, rollback availability,
and override options; T3 actions affecting identity, customer-facing assets, or cross-tenant
policy require dual approval. \emph{Override Latency} (OLA) and \emph{Override Viability Score}
(OVS) instantiate $P_{\mathrm{int}}$ of Eq.~\eqref{eq:pint} as operational evidence.

\paragraph{R3 -- Rollback and recovery.}
\emph{Engineering claim:} every reversible action ships with a rehearsed rollback; every
high-impact action ships with a documented minimum-risk fallback. Playbooks specify
preconditions, postconditions, dependencies, privileges, recovery order, and verification.
\emph{Recovery Completion Time} (RCT) measures rollback initiation to safe-state restoration; if
rehearsal fails or RCT exceeds budget, T2 or T3 authority is suspended until retested.

\paragraph{R4 -- Safe-state transition.}
\emph{Engineering claim:} a qualified operator must move the agent to $s_3$ through at least two
independent channels, and the transition must \emph{shrink} the agent's authority surface rather
than open new attack paths. Invocations are logged through the R1 channel; evidence includes
safe-state-transition logs, break-glass records, dependency maps, and post-halt review records.

\paragraph{AIBOM-linked traceability.}
Models, prompts, policies, plugins, guardrails, and playbooks bind to AIBOM identifiers so that
post-incident reconstruction identifies which upstream artifact influenced the action.

\medskip
\noindent R1--R4 turn \secodd into an auditable security-engineering framework: R1 preserves
\emph{what happened}, R2 whether a qualified human \emph{could intervene}, R3 whether the system
\emph{could recover}, and R4 whether autonomy \emph{could be stopped safely}.
Section~\ref{sec:scenarios} converts these into per-scenario pass/fail audit outcomes.

%% file: sec/6_scenarios_4.tex
\section{Scenario-Based Evaluation and Operational Mapping}
\label{sec:scenarios}
This section evaluates whether \secodd can structure high-impact ACD failures into reconstructable audit judgments. 
The evaluation is analytic and scenario-based, not a measurement of deployed SOC performance. 
We apply the same five-step walk to each scenario: control-state diagnosis, ODD conformance, \rade evidence review, \msh evaluation, and remediation. Table~\ref{tab:auditmatrix} summarizes the outcomes. ODD passes when the action-context tuple satisfies the declared $ODD_{sec}$ or $\Phi$; Evidence passes when $EC \geq 0.95$; HITL passes when $p95~OLA \leq \tau_O$ and T3 dual approval is recorded; and Recovery passes when rollback or bounded safe-state fallback is rehearsed. These thresholds are operator-configurable defaults, not universal constants. Accordingly, the checkmarks in Table~\ref{tab:auditmatrix} denote illustrative audit outcomes obtained by applying the \secodd logic to the scenario assumptions, rather than empirical measurements from a deployed SOC.

\begin{table}[t]
\centering
\caption{Illustrative Scenario-level audit outcomes. Criteria are defined in  Section~\ref{sec:scenarios}.; the Outcome column summarizes the accountability result structured by Eq.~\eqref{eq:Lfn}.}

\label{tab:auditmatrix}
\scriptsize
\setlength{\tabcolsep}{2pt}
\begin{tabularx}{\textwidth}{p{1.45cm}p{0.75cm}p{1.55cm}p{0.9cm}p{1.25cm}p{1.35cm}X}
\toprule
\textbf{Scenario} & \textbf{Tier} & \textbf{ODD} & \textbf{Evid.} &
\textbf{HITL} & \textbf{Recov.} & \textbf{Outcome} \tabularnewline
\midrule
A -- FP isolation &
T2 ($s_1$) &
$\times$ no change context &
$\checkmark$ EC &
$\checkmark$ OLA &
$\checkmark$ rollback &
Deployer context-sync gap; vendor drift testing; \msh suspended for affected playbook. \tabularnewline
\midrule
B -- mass lockout &
T3 ($s_2$) &
$\times$ maintenance drift &
$\checkmark$ EC &
$\times$ no pre-action approval &
$\times$ no bounded fallback &
T3 authority suspended; tenant-wide revocation removed from autonomous authority. \tabularnewline
\bottomrule
\end{tabularx}
\end{table}

\subsection{Scenario A -- Ransomware Auto-Isolation False Positive}
\label{sec:scenA}

A T2 agent observes mass file-renaming on a critical application server and executes network
isolation, session termination, and service-principal freeze. The activity is actually an
approved bulk migration, so the quarantine disrupts a customer-facing service. The risk class is well documented: security tooling and automated defensive actions can themselves become service-impacting failure sources when deployed without adequate context, validation, and rollback controls~\cite{ligo2022acdRisk,acd-isa2023}. The 2024 CrowdStrike outage is used only as external grounding for this broader service-impacting failure mode, not as an instance of ACD false-positive containment~\cite{microsoft2024crowdstrike}.

\emph{Walk.} The control state is $s_1$ but the declared ODD does not encode the migration
window, so the action is ODD-nonconforming. The \rade review attributes the gap to the deployer
(change-context sync), supplier (feed-drift sensitivity), and infrastructure provider
(evidence-grade log delivery). Table~\ref{tab:auditmatrix}, row~A, shows ODD nonconformance with
EC, OLA, and rollback all passing, so Eq.~\eqref{eq:Lfn} loads the $\alpha$ term against the
deployer while $\beta,\gamma$ remain bounded. \emph{Remediation:} downgrade to $s_0$ when change
context is missing, require two independent ransomware indicators before containment, and
suspend T2 authority after recurring false positives on the same critical asset class.

\subsection{Scenario B -- Mass Legitimate-Account Lockout}
\label{sec:scenB}

Scenario B assumes a misconfigured or over-authorized T3 deployment in which a credential-protection agent is permitted to invoke tenant-wide credential revocation.
After maintenance-induced clock drift, the agent observes a burst of failed TOTP authentications and misclassifies the pattern as credential stuffing. TOTP and HOTP correctness depend on time and counter synchronization~\cite{rfc6238,rfc4226},
so an authentication-failure burst can indicate either an attack or benign drift.

\emph{Walk.} The agent operated at $s_2$ because tenant-wide revocation is high-impact, but the maintenance-drift condition lies outside the declared ODD. The \rade review identifies a missing maintenance-mode signal, missing drift-scenario evaluation, delayed escalation, and no blast-radius cap. Table~\ref{tab:auditmatrix}, row~B, shows ODD nonconformance and HITL failure because the escalation occurred only after the T3 action; thus $\alpha$ and $\beta$ in Eq.~\eqref{eq:Lfn} are both positive and T3 authority is suspended.

\emph{Remediation:} remove tenant-wide credential revocation from default autonomous authority; require dual approval by a SOC supervisor and IAM owner for any T3 identity-plane action; and, if a time-sync anomaly or maintenance flag is active,
force transition to $s_0$ or $s_3$ and use bounded alternatives such as rate limiting, step-up authentication, source-scoped blocking, session challenge, or selective hold.
\subsection{Korean Operational-Profile Mapping} \label{sec:kr} 

Korea's AI Framework Act, KISA AI security guidance, and public-sector AI security guidebook~\cite{kr-aiact,kisa-ai-guideline,nis-ai-guidebook} provide principles for trustworthy AI and AI security, but do not yet specify fine-grained Security ODD declarations, control-state delegation, action-time evidence channels, or \msh-style due-care evidence for high-impact ACD. A compact operational mapping follows: \textbf{institutional}, a national or sectoral expert
body defines a \secodd-aligned profile and coordinates evidence reconstruction;
\textbf{technical}, an evidence profile codifies tamper-evident logging, version binding, override availability, and rollback rehearsal; \textbf{operational}, deployers classify ACD actions by tier, declare playbook-specific Security ODDs, and rehearse override and rollback. The mapping is operational, not legal.

%% file: sec/7_conclusion_3.tex
\section{Discussion and Conclusion}
\label{sec:conclusion}

Once an ACD agent affects production services or identity infrastructure, the governance
question is no longer \emph{whether} autonomy is acceptable in the abstract, but whether each
mitigation action is bounded, interruptible, recoverable, and reconstructable from evidence.
\secodd answers this through verifiable boundaries (Security ODD, $\Omega$, $P_{\mathrm{int}}$,
\rade, \msh, R1--R4) that support scenario-level pass/fail audit outcomes (Table~\ref{tab:auditmatrix}). The framework is scenario-analytic; the audit thresholds and the
weights in Eq.~\eqref{eq:Lfn} require sector-specific calibration against shadow-mode telemetry,
override drills, and incident-attestation packages from real SOC deployments, and the Korean
instantiation is operational, not legal. \secodd authorizes ACD only when execution-time control
is structured, intervention is meaningful, evidence is tamper-evidently preserved, and
post-incident responsibility can be reconstructed.